\documentstyle[12pt]{article}

\def\hybrid{\topmargin 0pt	\oddsidemargin 0pt
	\headheight 0pt	\headsep 0pt
	\textwidth 6.25in	
	\textheight 9.5in	
	\marginparwidth .875in
	\parskip 5pt plus 1pt	\jot = 1.5ex}

\hybrid

\catcode`\@=11

\def\marginnote#1{}
\newcount\hour
\newcount\minute
\newtoks\amorpm
\hour=\time\divide\hour by60
\minute=\time{\multiply\hour by60 \global\advance\minute by-\hour}
\edef\standardtime{{\ifnum\hour<12 \global\amorpm={am}%
	\else\global\amorpm={pm}\advance\hour by-12 \fi
	\ifnum\hour=0 \hour=12 \fi
	\number\hour:\ifnum\minute<10 0\fi\number\minute\the\amorpm}}
\edef\militarytime{\number\hour:\ifnum\minute<10 0\fi\number\minute}

\def\draftlabel#1{{\@bsphack\if@filesw {\let\thepage\relax
   \xdef\@gtempa{\write\@auxout{\string
      \newlabel{#1}{{\@currentlabel}{\thepage}}}}}\@gtempa
   \if@nobreak \ifvmode\nobreak\fi\fi\fi\@esphack}
	\gdef\@eqnlabel{#1}}
\def\@eqnlabel{}
\def\@vacuum{}
\def\draftmarginnote#1{\marginpar{\raggedright\scriptsize\tt#1}}

\def\draft{\oddsidemargin -.5truein
	\def\@oddfoot{\sl preliminary draft \hfil
	\rm\thepage\hfil\sl\today\quad\militarytime}
	\let\@evenfoot\@oddfoot	\overfullrule 3pt
	\let\label=\draftlabel
	\let\marginnote=\draftmarginnote
   \def\@eqnnum{(\theequation)\rlap{\kern\marginparsep\tt\@eqnlabel}%
\global\let\@eqnlabel\@vacuum}  }

\def\preprint{\twocolumn\sloppy\flushbottom\parindent 2em
	\leftmargini 2em\leftmarginv .5em\leftmarginvi .5em
	\oddsidemargin -.5in	\evensidemargin -.5in
	\columnsep .4in	\footheight 0pt
	\textwidth 10.in	\topmargin  -.4in
	\headheight 12pt \topskip .4in
	\textheight 6.9in \footskip 0pt
	\def\@oddhead{\thepage\hfil\addtocounter{page}{1}\thepage}
	\let\@evenhead\@oddhead	\def\@oddfoot{}	\def\@evenfoot{} }

\def\numberbysection{\@addtoreset{equation}{section}
	\def\theequation{\thesection.\arabic{equation}}}

\def\underline#1{\relax\ifmmode\@@underline#1\else
	$\@@underline{\hbox{#1}}$\relax\fi}

\def\titlepage{\@restonecolfalse\if@twocolumn\@restonecoltrue\onecolumn
     \else \newpage \fi \thispagestyle{empty}\c@page\z@
	\def\thefootnote{\fnsymbol{footnote}} }

\def\endtitlepage{\if@restonecol\twocolumn \else \newpage \fi
	\def\thefootnote{\arabic{footnote}}
	\setcounter{footnote}{0}}  

\catcode`@=12
\relax

\def\figcap{\section*{Figure Captions\markboth
	{FIGURECAPTIONS}{FIGURECAPTIONS}}\list
	{Figure \arabic{enumi}:\hfill}{\settowidth\labelwidth{Figure 999:}
	\leftmargin\labelwidth
	\advance\leftmargin\labelsep\usecounter{enumi}}}
 \relax
\def\tablecap{\section*{Table Captions\markboth
	{TABLECAPTIONS}{TABLECAPTIONS}}\list
	{Table \arabic{enumi}:\hfill}{\settowidth\labelwidth{Table 999:}
	\leftmargin\labelwidth
	\advance\leftmargin\labelsep\usecounter{enumi}}}
 \relax
\def\reflist{\section*{References\markboth
	{REFLIST}{REFLIST}}\list
	{[\arabic{enumi}]\hfill}{\settowidth\labelwidth{[999]}
	\leftmargin\labelwidth
	\advance\leftmargin\labelsep\usecounter{enumi}}}
 \relax
\makeatletter
\newcounter{pubctr}
\def\publist{\@ifnextchar[{\@publist}{\@@publist}}
\def\@publist[#1]{\list
	{[\arabic{pubctr}]\hfill}{\settowidth\labelwidth{[999]}
	\leftmargin\labelwidth
	\advance\leftmargin\labelsep
	\@nmbrlisttrue\def\@listctr{pubctr}
	\setcounter{pubctr}{#1}\addtocounter{pubctr}{-1}}}
\def\@@publist{\list
	{[\arabic{pubctr}]\hfill}{\settowidth\labelwidth{[999]}
	\leftmargin\labelwidth
	\advance\leftmargin\labelsep
	\@nmbrlisttrue\def\@listctr{pubctr}}}
 \relax
\makeatother
\newskip\humongous \humongous=0pt plus 1000pt minus 1000pt

\newif\ifdtup

\relax

\def\thefootnote{\fnsymbol{footnote}}
\def\ref#1{$^{#1)}$}
\def\p{\partial}
\def\pb{{\bar \partial}}

\def\ub{{\bar u}}
\def\e{E_{u}}
\def\eb{E_{\bar u}}
\def\x{\psi_{+}}
\def\xb{\psi_{-}}
\def\u{\underline}
\def\ds{\delta^{(s)}_{\varepsilon}}
\def\dd{\delta^{(2)}_{\varepsilon}}

\def\s{\sigma}
\def\ee{\varepsilon}
\begin{document}
\begin{titlepage}
\begin{center}
\hfill LPTENS-92-30\\
           \hfill October 1992\\
\hfill hep-th/9211083\\
\vskip 1in

{\large \bf Target Space Description of $W_{\infty}$ Symmetry in Coset Models}

\vskip .8in

\u{Ioannis Bakas}\footnote{Presently serving in the Greek Armed Forces.} \\
\vskip .1in
{\em Department of Physics \\
     University of Crete\\
     Heraklion, GR 71409, GREECE}
\vskip .4in

and
\vskip .4in

\u{Elias Kiritsis}\footnote{Present address: Theory
Division, CERN, CH-1211, Geneva 23, Switzerland, email:
kiritsis@surya1.cern.ch}
\vskip .1in

{\em  Laboratoire de Physique Th\'eorique\\
      de l'Ecole Normale Sup\'erieure\\
      24 rue Lhomond\\
      Paris, CEDEX 05, F-75231, FRANCE}
\end{center}
\vskip .8in

\begin{abstract}
We derive the transformation laws in the target space of the $SL(2,R)/U(1)$
coset model which generate the classical $W_{\infty}$ symmetry.
\end{abstract}
\end{titlepage}
\newpage
\renewcommand{\thepage}{\arabic{page}}
\setcounter{page}{1}
It was realized recently that the concept of $W$-algebras \cite{ZF}, their
decompactification limits (algebras of the $W_{\infty}$ type) \cite{B,PRS,BK}
and their non-compact analogues \cite{BK2}, play an important
role in such varied subjects as CFTs, their integrable off-critical siblings,
integrable hierarchies, string theory both perturbative and non-perturbative,
etc.
In this context, $W$-algebras are described as abstract operator algebras
that organize the operator content of 2-d theories, they are local and
generically
contain higher spin operators, which are responsible for their non-linearity.
They reflect the underlying parafermionic symmetry of CFTs,
which is generically non-local, and can be simply described  as a local
subalgebra of the enveloping
parafermionic algebra.

In the $\s$-model description of such CFTs, the presence of the $W$-algebras
is associated with the existence of chiral symmetries.
However, in the $\s$-model context one usually works in perturbation
theory around the decompactification limit ($\alpha '$ $\rightarrow 0$).
Thus, classically, the symmetry will be of the $W_{\infty}$ type, and
$\alpha '$ corrections will renormalize it accordingly; for example, in the
compact case the renormalized $W$-algebra should be finitely generated.
Such a target space description of $W$-symmetries has not attracted any
attention so far.
We feel that it is interesting to explore it, in order to get a better grasp
of $W$-symmetries from a (more conventional) lagrangian point of view, which
usually provides an easier link to geometric concepts and their
generalizations.

The purpose of this letter is to use the parafermionic description of
$W$-algebras together with the $\s$-model description of the parafermionic
currents \cite{BCR} in order to obtain a systematic formulation of
$W$-symmetries in target space.
The CFT models we will be considering are of the coset ($G/H$) type for which
a $\s$-model description is available as gauged WZW models \cite{gWZW}.
Our method is applicable to all such models, both compact and non-compact,
but for notational simplicity we will only consider the $SL(2,R)/U(1)$ model.
At the clasical level the $W$-symmetry is identical for various non-compact
versions of a compact target space (that is the manifolds obtained by analytic
continuation of the compact manifold), for example the $N\rightarrow \infty$
limit of the $W_{N}$ algebra associated to the $SU(2)_{N}/U(1)$ model
and the $k\rightarrow \infty$ limit of the ${\hat W}_{\infty}(k)$ algebra of
the $SL(2,R)_{k}/U(1)$ model coincide \cite{BK2}.

The $SL(2,R)_{k}/U(1)$ coset model has received considerable attention
recently since it admits a geometric interpretation (for $k$ large) as a 2-d
black hole \cite{Wit}.
Writing down the gauged WZW action and integrating out the gauge fields we
obtain to leading order in $k$ the following action (in isothermal coordinates)
$$S={k\over 4\pi}\int {\p u\pb \ub +\p \ub \pb u \over 1-u\ub}.\eqno(1)$$
It differs from the conventional $\s$-model on $S^{2}$ (or its compact
analogue) in that the
target space metric is $(1-u\ub)^{-1}$ instead of $(1-u\ub)^{-2}$.
Also, the difference between $SU(2)/U(1)$ and $SL(2,R)/U(1)$ is reflected in
the range of the coordinates $u,\ub$.
The classical equations of motion can be derived from the variation of $S$
$$\delta S\sim \int (\delta u \eb +\delta \ub \e).\eqno(2)$$
They read
$$\e=\eb=0\eqno(3)$$
with
$$\e ={\p\pb u\over 1-u \ub} +{\ub \p u \pb u \over
(1-u\ub)^{2}}\,\,\,,\eqno(4a)$$
$$\eb={\p\pb\ub\over 1-u\ub}+{u\p\ub\pb\ub\over (1-u\ub)^{2}}\,\,\,.\eqno(4b)$$

The action (1) has a classical $U(1)$ symmetry
$$u\rightarrow ue^{i\alpha}\,\,\,,\,\,\, \ub\rightarrow \ub
e^{-i\alpha}\eqno(5)$$
reflecting the killing symmetry of the the target manifold. The associated
current is
$$J={\ub\p u-u\p\ub\over 1-u\ub}\,\,\,,\,\,\,{\bar J}={\ub\pb u-u\pb\ub\over
1-u\ub}\eqno(6)$$
which is conserved but not chiral
$$\p {\bar J}+\pb J=0\eqno(7)$$
using the equations of motion.

We can define the generating parafermion (non-local) currents in the
standard way \cite{BCR} by dressing the gauge currents with Wilson lines
\footnote{From now on we focus on the holomorphic part of the theory to
avoid repetition.}
$$\x={\p u \over \sqrt{1-u\ub}}V_{+}\,\,\,,\,\,\,\xb={\p\ub \over
\sqrt{1-u\ub}}V_{-}\eqno(8)$$
where
$$V_{\pm}=exp\left[\pm {1\over 2}\int_{C_{z}}(dzA+d{\bar z}{\bar A})\right]
\eqno(9)$$
$$A=-J\,\,\,,\,\,\,{\bar A}={\bar J}\eqno(10)$$
and $C_{z}$ is a path (string) terminating at the position of the parafermion.
The curvature of the gauge connection (10) is proportional to the equations
of motion
$$F(A)\equiv \p {\bar A}-\pb A =2(\ub\e -u\eb)\eqno(11)$$
and thus vanishes on-shell.
Consequently, the parafermion currents are insensitive to the position of
the string, ie., variations of the path $C_{z}$.
Straightforward computation gives
$$\pb \x =\sqrt{1-u\ub}\e V_{+}\;\;\;,\eqno(12a)$$
$$\pb \xb=\sqrt{1-u\ub}\eb V_{-}\eqno(12b)$$
which implies the chiral conservation of parafermion currents on-shell.
Any chirally conserved field in the theory is a polynomial in $\x$, $\xb$.
However of primary interest are fields which are free of the string
non-locality.
Such fields are the W-currents and their descendants.
A convenient (quasiprimary) basis of such currents, up to an overall
spin-dependent normalization, is \cite{BK}
$$W_{s}=\sum_{k=1}^{s-1}(-1)^{s-k-1}A^{s}_{k}\p^{k-1}\x\p^{s-k-1}\xb\eqno(13a)$$
with spin $s=2,3,...$ and
$$A^{s}_{k}={1\over s-1}{s-1\choose k}{s-1\choose s-k}\;\;.\eqno(13b)$$
It is obvious that $W_{s}$ is a local functional of $u,\ub$ and their
derivatives, and any other local current is a polynomial of the $W's$
and their derivatives.

Having the W-currents, we can determine the symmetries responsible for their
conservation.
In particular we have to determine the variations $\ds u$, $\ds \ub$
such that
$$\delta S \sim \int (\ds u \eb +\ds \ub \e)=\int \varepsilon \pb W_{s}
\eqno(14)$$
where $\ee$ is a function of $z$ only.
The strategy is to start from the right-hand side of eq. (14) which is known,
bring it into the form dictated by the left-hand side and then simply read
the $u,\ub$ variations.

In the simplest case, s=2, using eqs. (12), (13), we have
$$\int \varepsilon \pb (\x\xb)=\int (\varepsilon\p u \eb +
\varepsilon\p \ub \e)\eqno(15)$$
which in turn implies
$$\dd u =\ee \p u \,\,\,,\,\,\,\dd \ub =\ee \p \ub\eqno(16)$$
as expected for the transformation laws of scalars that generate the stress
tensor of the theory.
We can easily verify on $u,\ub$
$$\left[ \delta^{(2)}_{\ee_{1}}, \delta^{(2)}_{\ee_{2}}\right]=
\delta^{(2)}_{\ee_{1}'\ee_{2}-\ee_{1}\ee_{2}'}\eqno(17)$$
which gives the centerless Virasoro algebra; here prime stands for the
derivative with respect to $z$.

The computation of the higher spin variations of $u,\ub$ proceeds
along similar lines.
The only point one has to be careful about is the non-commutativity of
the derivatives $\p$, $\pb$ on $\psi_{\pm}$ off-shell.
In fact, the non-vanishing curvature (11) yields
$$[\p,\pb]\psi_{\pm}=\pm (\ub \e-u\eb)\psi_{\pm}\eqno(18)$$
and more generally
$$[\p^{n}, \pb]\psi_{\pm}=\pm\sum_{l=0}^{n-1}{n\choose l}\p^{n-l-1}
(\ub\e-u\eb)\p^{l}\psi_{\pm}\;\;\;.\eqno(19)$$

The results for the spin 3 and 4 transformations are
$$\delta^{(3)}_{\ee}u=2\ee\left(\p^{2}u+2{u\p u\p\ub\over 1-u\ub}\right)+
\ee'\p u\;\;\;,\eqno(20a)$$
$$\delta^{(3)}_{\ee}\ub=-2\ee\left(\p^2 \ub+2{\ub\p u\p\ub\over 1-u\ub}
\right)-\ee'\p\ub\eqno(20b)$$
and
$$\delta^{(4)}_{\ee}u=5\ee\left(\p^3 u+{\p\ub(2(\p u)^{2}+3u\p^2 u)\over
1-u\ub}+3{u^{2}\p u(\p \ub)^{2}\over (1-u\ub)^{2}}\right)+$$
$$+5\ee'\left(\p^2 u+{u\p u\p\ub\over 1-u\ub}\right)+\ee''\p
u\;\;\;,\eqno(21a)$$
$$\delta^{(4)}_{\ee}\ub=5\ee\left(\p^3\ub+{\p u(2(\p\ub)^{2}+3\ub\p^2\ub)
\over 1-u\ub}+3{\ub^{2}\p\ub(\p u)^{2}\over (1-u\ub)^{2}}\right)+$$
$$+5\ee'\left(\p^2\ub+{\ub\p u\p\ub\over
1-u\ub}\right)+\ee''\p\ub\;\;\;.\eqno(21b)$$

The transformation for arbitrary $s$ is
$$\ds u=\sum_{l=0}^{s-2}B^{s}_{l}\p^l \ee{1-u\ub\over \p\ub}
\psi_{-}\p^{s-l-2}\psi_{+}+$$
$$+\sum_{k=1}^{s-1}\sum_{l=0}^{k-2}(-1)^{l}A^{s}_{k}{k-1\choose l}
u\p^{k-2-l}\left[\ee\p^{l}\psi_{-}\p^{s-k-1}\psi_{+}-(-1)^s\ee\p^{l}
\psi_{+}\p^{s-k-1}\psi_{-}\right]\eqno(22a)$$
with
$$B^{s}_{l}={1\over s-1}{2s-l-2\choose s}{s-1\choose l}\;\;\;.\eqno(22b)$$
The variation of $\ub$ is also given by eq. (22) by interchanging
$u\leftrightarrow \ub$ and multiplying by a factor $(-1)^{s}$.
The explicit dependence of these transformations on $u,\ub$ is obtained using
the formulae
$$\p^{n}\psi_{+}=V_{+}\left(\p+{1\over 2}A\right)^{n}{\p u\over \sqrt{
1-u\ub}}\;\;\;,\eqno(23a)$$
$$\p^{n}\psi_{-}=V_{-}\left(\p-{1\over 2}A\right)^{n}{\p \ub\over \sqrt{
1-u\ub}}\;\;\;\eqno(23b)$$
where $(\p\pm A/2)^{n}$ acts as an operator on the right.

Commutators of the variations $\ds$ on $u,\ub$ give rise to the
centerless $W_{\infty}$
algebra, \cite{PRS,BK}, as expected.
In particular, apart from the commutation relations (17), we have after
rescaling $\delta^{(s)}_{\varepsilon}$ by $2^{s-3}s!/(2s-3)!!$ that
$$[\delta^{(2)}_{\ee_{1}},\delta^{(3)}_{\ee_{2}}]=\delta^{(3)}_{2\ee_{1}'
\ee_{2}-\ee_{1}\ee_{2}'}\;\;\;,\eqno(24)$$
$$[\delta^{(2)}_{\ee_{1}},\delta^{(4)}_{\ee_{2}}]=\delta^{(4)}_{3\ee_{1}'
\ee_{2}-\ee_{1}\ee_{2}'}+{32\over 5}\delta^{(2)}_{\ee_{1}'''\ee_{2}}\;\;\;,
\eqno(25)$$
$$[\delta^{(3)}_{\ee_{1}},\delta^{(3)}_{\ee_{2}}]=2\delta^{(4)}_{\ee_{1}'
\ee_{2}-\ee_{1}\ee_{2}'}+{4\over 5}
\delta^{(2)}_{2(\ee_{1}'''\ee_{2}-\ee_{1}\ee_{2}''')+
3(\ee_{1}'\ee_{2}''-\ee_{1}''\ee_{2}')}\eqno(26)$$
and in general
$$[\delta^{(s)}_{\ee_{1}},\delta^{(s')}_{\ee_{2}}]=
\delta^{(s+s'-2)}_{(s'-1)\ee_{1}'\ee_{2}-
(s-1)\ee_{1}\ee_{2}'}+{\rm lower}\,\,\,{\rm spin}\,\,\,{\rm terms}.\eqno(27)$$
These commutation relations can be compared with  the OPE of the $W_{\infty}$
algebra using the correspondence
$$\delta^{(s)}_{\ee}\rightarrow \oint \ee(z)W^{s}(z)\;\;\;.\eqno(28)$$

One should note that the behaviour of W-currents and transformations
is special for target space ``instantons".
Here by instantons we simply mean holomorphic maps, $u(z)$, $\ub({\bar z})$.
They are not exactly instantons since the would be topological charge
being proportional to the volume is infinite.
Holomorphic maps are obviously solutions to the equations of motion (3),(4).
The W-currents vanish when evaluated on holomorphic maps.
As for the W-transformations they transform $u$ linearly and holomorphically
but not $\ub$.
Their explicit form is
$$\ds u =\sum_{l=0}^{s-2}B^{s}_{l}\p^{l}\ee
\p^{s-l-1}u\;\;\;,\;\;\;\delta^{(s)}
_{\varepsilon}\ub=0\eqno(29)$$
where $B^{s}_{l}$ has been defined in eq. (22b).
They are the same as the W-transformations in flat space \cite{BK}.
The situation has similarities with the geometric picture of W-algebras
advanced in \cite{GM}.
One should also note that the $W_{\infty}$ transformations (29) generate a
group which
is the Borel type (differential) subgroup of the Lie-Poisson group of
pseudodifferential operators.
It is interesting that in the case of non-holomorphic maps one has a
generalization of this structure which couples $u$ and $\ub$.

Once we go to next order (one loop) in $1/N$ or $1/k$, the effects of the
measure
(dilaton) become important. Although at the level of the $\s$-model
it is difficult to analyze the renormalization of the W-symmetry, we know
its fate from the operator approach \cite{ZF,BK2}.
In the compact case there are renormalizations which render the W-generators
with $s>N$ null and thus they decouple, \cite{BK2}.
This does not happen in the non-compact, SL(2,R), case. It would be interesting
to verify this also via a (perturbative) $\sigma$-model calculation.
Also, the fate of the $U(1)$ symmetry
is different in the compact or non-compact case.
In the compact case it is broken down to the $Z_{N}$ parafermionic symmetry,
whereas in the non-compact case it is unbroken and provides the
``time"-translation symmetry in the black
hole background, \cite{BK2}. In both cases the U(1) current is absent from
the Hilbert space due to IR divergences.
The breaking of the $U(1)$ symmetry to $Z_{N}$ is reminiscent of the analogous
phenomenon in QCD: the breaking of the axial $U(1)$ symmetry to $Z_{N_{f}}$,
by instantons. In QCD this has as a result that certain multifermion operators
with non-zero axial charge to get an expectation value. The same is also true
in the parafermion model where now the role of the fermions is played by the
generating parafermions.

It would be interesting to understand further the target space structure
of the W-symmetries.
Also, it would be interesting to understand the relation (if any) between
$W_{\infty}$ and more generally W-symmetries of CFT coset models and the hidden
symmetries (of the Kac-Moody type) of ordinary $\sigma$-models, \cite{wu}
both from the algebraic and langrangian point of view.
Work in this direction is in progress.

\vskip 1in
\centerline {Acknowledgements}
\smallskip
Part of this work was done while one of us (I. B.) was a visiting member
at the IAS at Princeton in the fall of 1991.
E.K. would like to thank J. L. Gervais for a discussion.

\end{document}